\documentstyle[twocolumn,pra,aps,psfig]{revtex}

\def\Im{\mathop{\rm Im}}

\begin{document}
\title{Quantum fluctuations of position of a mirror in vacuum}
\author{Marc Thierry Jaekel $^{(a)}$ and Serge Reynaud $^{(b)}$}
\address{(a) Laboratoire de Physique Th\'{e}orique de l'ENS
\thanks{%
Unit\'e propre du Centre National de la Recherche Scientifique,
associ\'ee \`a l'Ecole Normale Sup\'erieure et \`a l'Universit\'e
Paris-Sud}, 24 rue Lhomond, F75231 Paris Cedex 05 France\\
(b) Laboratoire de Spectroscopie Hertzienne
\thanks{%
Unit\'e de l'Ecole Normale Sup\'erieure et de l'Universit\'e
Pierre et Marie Curie, associ\'ee au Centre National de la Recherche 
Scientifique}, 4 place Jussieu, case 74, F75252 Paris Cedex 05 France}
\date{{\sc Journal de Physique I} {\bf 3} (1993) 1-20}
\maketitle

\begin{abstract}
A mirror scattering vacuum fields is submitted to a quantum fluctuating
radiation pressure. It also experiences a motional force, related to force
fluctuations through fluctuation-dissipation relations. The resulting
position fluctuations of the coupled mirror are related to the dissipative
part of the mechanical admittance. We compute the time dependent position
commutator, which makes apparent the difference between the low-frequency
and high-frequency masses, and the anticommutator noise spectrum, which
describes the ultimate sensitivity in a length measurement using mirrors.

\end{abstract}

\section*{Introduction}

A mirror which scatters a laser field is submitted to a fluctuating
radiation pressure \cite{Position1}, which is associated through
fluctuation-dissipation relations with a motional force, that is a mean
force for a moving mirror \cite{Position2}. The resulting random motion has
been studied in detail since it determines the ultimate sensitivity of the
interferometers designed for gravitational wave detection \cite{Position3}.

In the absence of laser irradiation, the mirror still scatters vacuum
fields, so that the radiation pressure fluctuates \cite{Position4}, and the
motional force does not vanish \cite{Position5,Position6}. Fluctuations and
dissipation are also directly connected in this case \cite{Position7}. As a
consequence of its coupling with vacuum radiation pressure, the mirror's
position has to acquire some quantum fluctuations, even if it is introduced
as a classical variable. On the other hand, the mirror has proper quantum
fluctuations, even if it is not coupled to the field. The purpose of the
present paper is to elucidate the relation between those acquired and proper
fluctuations.

In order to reach consistent results, it is necessary that the mirror behave
as a stable system when coupled to vacuum. For a perfect mirror however,
`runaway solutions' appear, which are analogous to the ones encountered in
classical electron theory \cite{Position8}. We will use the fact that this
instability problem is solved \cite{Position9} by considering a partially
transmitting mirror, causal and transparent at frequencies higher than a
reflection cutoff $\omega _{C}$ much smaller in reduced units than the
mirror's mass $m_{0}$ 
\begin{equation}
\hbar \omega _{C}\ll m_{0}c^{2}  \eqnum{1}
\end{equation}
This condition, which allows one to ignore the recoil effect in the
derivation of the motional force, also implies passivity, that is the
incapacity of vacuum to sustain runaway solutions. Then, stability results
from passivity \cite{Position9}.

We shall obtain that the position fluctuations of the coupled mirror are
connected to the dissipative part of the mechanical admittance, in analogy
with the thermalized position of an atomic system scattering a thermal field 
\cite{Position10}, or the thermalized voltage of an electrical oscillator
coupled to a resistor at thermal equilibrium \cite{Position11}. However, at
zero temperature, this equilibrium involves vacuum fluctuations rather than
thermal ones \cite{Position12}. The non-commutativity of the force
correlation functions will play a central role in this paper, as well as the
relation of the commutator with the dissipative part of the motional
susceptibility \cite{Position13}.

First, we study the `input fluctuations', i.e. the fluctuations when
coupling is disregarded. The mirror's dynamics are described by two noise
spectra corresponding respectively to the commutator and anticommutator,
which are related to a mechanical admittance function and which behave
resonantly in the vicinity of suspension eigenfrequencies. When coupling to
vacuum radiation pressure is taken into consideration, the admittance
function is modified. Using the techniques of linear response theory \cite
{Position13}, we deduce coupled fluctuations which are related to the
modified admittance through fluctuation-dissipation relations. In this
derivation, input position fluctuations are considered on an equal foot with
input force fluctuations. Yet, it appears that the fluctuations of the
coupled variables are completely determined by those of the input force.
This means that the results might be obtained from a Langevin equation \cite
{Position14}, where all fluctuations are fed by the input force
fluctuations; this simpler derivation is presented in Appendix A.

Then, we discuss the two noise spectra related respectively to the
commutator and to the anticommutator for the coupled position. The former
describes a modification of the canonical commutation relation. Its time
dependence, analyzed in Appendix B, appears to be connected to a difference
between the low-frequency and high-frequency values of mass. The latter
gives the quantum noise on the position of the coupled mirror. When
integrated over frequency, it provides the position variance; as a function
of time, it describes the quantum diffusion of a mirror coupled to vacuum
radiation pressure. The anticommutator noise spectrum is involved in
ultimate quantum limits for a position measurement \cite{Position3}.

We also analyse the autocorrelation of the coupled force and its cross
correlation with the coupled position.

For the sake of simplicity, our attention will focus upon the simple problem
of an harmonically bound mirror. However, linear response formalism can also
be used for the anharmonic oscillator, as shown in Appendix C.

\section*{Input position fluctuations}

In a stationary state, we will write any correlation function as the sum of
an antisymmetric part, associated with a commutator, and of a symmetric one,
related to an anticommutator 
\begin{eqnarray}
C_{AA}(t) &=&\left\langle A(t)A(0)\right\rangle -\left\langle A\right\rangle
^{2}=\hbar \left( \sigma _{AA}(t)+\xi _{AA}(t)\right)  \eqnum{2a} 
\end{eqnarray}
\begin{eqnarray}
C_{AA}(t)-C_{AA}(-t) &=&\left\langle A(t)A(0)-A(0)A(t)\right\rangle 
\nonumber \\
=2\hbar \xi _{AA}(t)  \eqnum{2b} \\
C_{AA}(t)+C_{AA}(-t) &=&\left\langle A(t)A(0)-A(0)A(t)\right\rangle
-2\left\langle A\right\rangle ^{2}
\nonumber \\
=2\hbar \sigma _{AA}(t)  \eqnum{2c}
\end{eqnarray}
The antisymmetric part $\xi _{AA}$, typical of quantum fluctuations, would
vanish for any classical stochastic treatment.

We will denote for any function $f$ 
\[
f(t)={\int }\frac{{\rm d}\omega }{2\pi }f[\omega ]e^{-i\omega t} 
\]
The function $\xi _{AA}[\omega ]$ is an odd function of $\omega $ while the
noise spectrum $\sigma _{AA}[\omega ]$ is an even one. If $A$ is a hermitean
operator, both are real functions of $\omega $.

Let us first express the position fluctuations of an harmonic oscillator
(mass $m_{0}$, eigenfrequency $\omega _{0}$) in its ground state within the
linear response formalism. The correlation functions of this `input
position' $q^{\rm in}$ are easily computed (we use shorthand notation where $\xi
_{qq}^{\rm in}$ stands for $\xi _{q^{\rm in}q^{\rm in}}$) 
\begin{eqnarray}
C_{qq}^{\rm in}(t) &=&\frac{\hbar }{2m_{0}\omega _{0}}\exp (-i\omega _{0}t) 
\eqnum{3a} \\
\xi _{qq}^{\rm in}(t) &=&\frac{-i}{2m_{0}\omega _{0}}\sin (\omega _{0}t) 
\eqnum{3b} \\
\sigma _{qq}^{\rm in}(t) &=&\frac{1}{2m_{0}\omega _{0}}\cos (\omega _{0}t) 
\eqnum{3c}
\end{eqnarray}
as well as the associated spectra 
\begin{eqnarray}
C_{qq}^{\rm in}[\omega ] &=&\frac{\hbar \pi }{m_{0}\omega _{0}}\delta (\omega
-\omega _{0})  \eqnum{3d} \\
\xi _{qq}^{\rm in}[\omega ] &=&\frac{\pi }{2m_{0}\omega _{0}}\left( \delta
(\omega -\omega _{0})-\delta (\omega +\omega _{0})\right)  \eqnum{3e} \\
\sigma _{qq}^{\rm in}[\omega ] &=&\frac{\pi }{2m_{0}\omega _{0}}\left( \delta
(\omega -\omega _{0})+\delta (\omega +\omega _{0})\right)  \eqnum{3f}
\end{eqnarray}
One also defines a mechanical susceptibility $\chi _{qq}^{\rm in}$ which
describes the linear response of the position to an exerted force \cite
{Position13} 
\begin{equation}
\chi _{qq}^{\rm in}[\omega ]=\frac{1}{m_{0}\left( \omega _{0}^{2}-(\omega
+i\epsilon )^{2}\right) }  \eqnum{3g}
\end{equation}
The parameter $\epsilon \rightarrow 0^{+}$ is inserted in order to ensure
the causal character of the susceptibility. The susceptibility is directly
connected to the mechanical impedance $Z^{\rm in}$ or to the mechanical
admittance $Y^{\rm in}$ 
\begin{equation}
-i\omega \chi _{qq}^{\rm in}[\omega ]=Y^{\rm in}[\omega ]=\frac{1}{Z^{\rm in}[\omega ]} 
\eqnum{4}
\end{equation}
Those functions obey the fluctuation-dissipation relations \cite{Position13}
applied to a quantum system at zero temperature 
\begin{eqnarray}
C_{qq}^{\rm in}[\omega ] &=&2\hbar \theta (\omega )\xi _{qq}^{\rm in}[\omega ] 
\eqnum{5a} \\
\sigma _{qq}^{\rm in}[\omega ] &=&\varepsilon (\omega )\xi _{qq}^{\rm in}[\omega
]\qquad \varepsilon (\omega )=\theta (\omega )-\theta (-\omega )  \eqnum{5b}
\\
2i\xi _{qq}^{\rm in}[\omega ] &=&\chi _{qq}^{\rm in}[\omega ]-\chi
_{qq}^{\rm in}[-\omega ]=2i\ \Im \chi _{qq}^{\rm in}[\omega ]  \eqnum{5c}
\end{eqnarray}
The $\theta (\omega )$ function may be considered as the limit at zero
temperature of Planck spectrum \cite{Position15} 
\[
\lim_{T\rightarrow 0}\frac{1}{1-\exp \frac{-\hbar \omega }{k_{B}T}}=\theta
(\omega ) 
\]

Quantities evaluated from the correlation functions at equal times, like the
canonical commutator or the variances, may be expressed as integrals over
frequency of the spectra $\xi _{qq}[\omega ]$ and $\sigma _{qq}[\omega ]$.
For instance, the commutator of two positions evaluated at different times 
\[
\left\langle \left[ q(t),q(0)\right] \right\rangle =2\hbar \xi _{qq}(t) 
\]
is a real and odd function of $t$. The commutator between velocity and
position is simply deduced ($v(t)$ is the mirror's velocity $q^{\prime }(t)$) 
\begin{equation}
\left\langle \left[ v(t),q(0)\right] \right\rangle =2\hbar \xi
_{vq}(t)=2\hbar \xi _{qq}^{\prime }(t)  \eqnum{6}
\end{equation}
Introducing a specific notation for integrals over frequency 
\begin{equation}
\overline{g}={\int }\frac{{\rm d}\omega }{2\pi }g[\omega ]=g(0)  \eqnum{7}
\end{equation}
the equal-time canonical commutator is 
\begin{equation}
\left\langle \left[ v(0),q(0)\right] \right\rangle =2\hbar \xi _{qq}^{\prime
}(0)=-2i\hbar \overline{\omega \xi _{qq}}  \eqnum{8}
\end{equation}
The symmetrical correlation function (we assume that the mean value 
$\left\langle q\right\rangle $ vanishes) 
\[
\left\langle q(t)q(0)+q(0)q(t)\right\rangle =2\hbar \sigma _{qq}(t) 
\]
is a real and even function of $t$, and the stationary variances are 
\begin{eqnarray}
\Delta q^{2} &=&\left\langle q(0)^{2}\right\rangle =\hbar \overline{\sigma
_{qq}}  \eqnum{9a} \\
\Delta v^{2} &=&\left\langle v(0)^{2}\right\rangle =\hbar \overline{\omega
^{2}\sigma _{qq}}  \eqnum{9b}
\end{eqnarray}
Using equations (3), one recovers the well known expressions 
\begin{eqnarray}
&&\left\langle \left[ v(0),q(0)\right] \right\rangle =\frac{-i\hbar }{m_{0}}
\eqnum{10} \\
\Delta q^{2} &=&\frac{\hbar }{2m_{0}\omega _{0}}\qquad \Delta v^{2}=\frac{
\hbar \omega _{0}}{2m_{0}}  \eqnum{11}
\end{eqnarray}
The covariance between position and velocity vanishes because $\sigma
_{qq}[\omega ]$ is an even function of $\omega $.

The foregoing description of quantum fluctuations may seem too sophisticated
for such a simple system as an harmonic oscillator. We see in the following
that this formalism is well suited to the analysis of fluctuations for the
system coupled to vacuum. Relations (5,6,8,9) will still be valid, with
different expressions for the noise spectra, and will have a significant
content in this case.

Furthermore, expressions (5) for the spectra contain more information than
the integrated quantities, even for the uncoupled variables. Consider for
instance that position is measured and that the measurement frequency band
is characterized by a function $G[\omega ]$ with a maximum value of 1 at the
frequency of the expected signal. Assuming that quantum noise may be dealed
with in the same manner as thermal noise, one gets the effective noise as an
integral over the detection bandwidth (with the notation of eq. 7) 
\[
\Delta q_{G}^{2}=\hbar \overline{G\sigma _{qq}} 
\]
This effective noise $\Delta q_{G}^{2}$ may be smaller than the position
variance $\Delta q^{2}$. It even vanishes if the suspension eigenfrequencies
are outside the detection bandwidth, which is the case in any high
sensitivity measurement; in the opposite case, the noise would be of the
order of the position variance $\Delta q^{2}$. This explains why the proper
quantum fluctuations of the mirror's position may be neglected when
analyzing the limits in position measurements. At the end of the paper, we
will come back to this discussion and show that the dissipative coupling
between mirror and vacuum sets an ultimate quantum limit in position
measurements.

\section*{Input force fluctuations}

We now consider the input force fluctuations $F^{\rm in}$, that is the
fluctuations of vacuum radiation pressure computed when disregarding
reaction of the mirror's motion.

These fluctuations are characterized by stationary correlation functions
denoted $C_{FF}^{\rm in}$, $\xi _{FF}^{\rm in}$ and $\sigma _{FF}^{\rm in}$, which are
connected through fluctuation-dissipation relations to the linear
susceptibility $\chi _{FF}^{\rm in}$ describing the motional force for small
displacements 
\begin{eqnarray}
C_{FF}^{\rm in}[\omega ] &=&2\hbar \theta (\omega )\xi _{FF}^{\rm in}[\omega ] 
\eqnum{12a} \\
\sigma _{FF}^{\rm in}[\omega ] &=&\varepsilon (\omega )\xi _{FF}^{\rm in}[\omega ] 
\eqnum{12b} \\
2i\xi _{FF}^{\rm in}[\omega ] &=&\chi _{FF}^{\rm in}[\omega ]-\chi
_{FF}^{\rm in}[-\omega ]=2i\ \Im \chi _{FF}^{\rm in}[\omega ]  \eqnum{12c}
\end{eqnarray}
These relations, which can be regarded as consequences of linear response
theory, have been directly checked for the problem of a mirror coupled to
vacuum radiation pressure \cite{Position7,Position9}. In particular,
equations (12a,b) correspond to the fact that vacuum is the equilibrium
state at zero temperature, so that radiation pressure can damp the mirror's
motion, but cannot excite it.

For a perfect mirror scattering a scalar vacuum field in a two-dimensional
(2D) spacetime, the motional force is proportional to the third derivative
of the position and corresponds to the following susceptibility, obtained by
linearizing the expressions obtained for a perfect mirror in reference \cite
{Position5}, or by considering the limit of perfect reflection in reference 
\cite{Position7} 
\[
\chi _{FF}^{\rm in}[\omega ]=\frac{i\hbar \omega ^{3}}{6\pi c^{2}}=im_{0}\tau
\omega ^{3} 
\]
The time constant $\tau $ characterizes the weak coupling of mirror to
vacuum 
\[
\tau =\frac{\hbar }{6\pi m_{0}c^{2}}
\]

A partially transmitting mirror \cite{Position7} is described by causal
reflectivity and transmittivity functions with a reflection cutoff. This
approach finds many justifications. It leads to an expression for the mean
Casimir force between two mirrors, which is free from the divergences
usually associated with the infiniteness of vacuum energy \cite{Position16}.
It also provides finite and causal expressions for the forces associated
with motions of two mirrors \cite{Position17}.

When condition (1) is obeyed, one can ignore the recoil effect in the
derivation of the motional force and obtain the susceptibility as 
\begin{equation}
\chi _{FF}^{\rm in}[\omega ]=im_{0}\tau \omega ^{3}\Gamma [\omega ] 
\eqnum{13a} 
\end{equation}
where $\Gamma [\omega ]$ is a cutoff function, whose expression is analysed
in detail in reference \cite{Position9}. A simple example, where the
aforementioned conditions are fulfilled, is provided by the following
lorentzian model for the reflectivity ($r$) and transmittivity ($s$)
functions 
\begin{eqnarray*}
r[\omega ] &=&\frac{-1}{1-\frac{i\omega }{\Omega }}\qquad s[\omega
]=1+r[\omega ] \\
\chi _{FF}^{\rm in}[\omega ] &=&6m_{0}\tau \Omega ^{3} \times
\nonumber \\
&&\left( -\frac{i\omega }
{\Omega }-\frac{\omega ^{2}}{2\Omega ^{2}}-\left( 1-\frac{i\omega }{\Omega }
\right) \ln \left( 1-\frac{i\omega }{\Omega }\right) \right)
\end{eqnarray*}
In the present paper, we will use only some general properties obeyed by the
cutoff function and not its explicit expression.

The function $\Gamma [\omega ]$ cuts off the high frequencies and is regular
at the low frequency limit 
\[
\Gamma _{0}=\Gamma [0]=r_{0}^{2} 
\]
$r_{0}$ is the reflection amplitude at $\omega =0$; $\Gamma _{0}=1$ for a
perfectly reflecting mirror at zero frequency. It can be written as the sum
of a dissipative and a dispersive parts 
\begin{eqnarray}
\Gamma [\omega ] &=&\Gamma _{R}[\omega ]+i\Gamma _{I}[\omega ]\qquad \Gamma
_{R}[\omega ]\ {\rm and}\ \Gamma _{I}[\omega ]\ {\rm real}  \nonumber \\
&&\xi _{FF}^{\rm in}[\omega ]=m_{0}\tau \omega ^{3}\Gamma _{R}[\omega ] 
\eqnum{13b}
\end{eqnarray}
The dissipative part $\Gamma _{R}$ obeys a positivity property \cite
{Position9} 
\begin{equation}
\Gamma _{R}[\omega ]\geq 0  \eqnum{14}
\end{equation}
The high frequency behaviour of $\Gamma _{R}[\omega ]$ will be important in
the following. Various conditions will be expressed as inequalities for a
positive coefficient $A$ 
\begin{equation}
A=-\lim_{\omega \rightarrow \infty }\frac{{\rm d}\ln \Gamma _{R}}{{\rm d}\ln
\omega }  \eqnum{15}
\end{equation}
Using the causal properties of the susceptibility, one can write a
dispersion relation with subtractions since the susceptibility does not
vanish at high frequencies. As the three quantities $\chi _{FF}^{\rm in}[0]$, 
$\chi _{FF}^{\rm in}{}^{\prime }[0]$ and $\chi _{FF}^{\rm in}{}^{\prime \prime }[0]$
vanish, this relation has a simple form in terms of the cutoff function \cite
{Position9} 
\[
\Gamma [\omega ]={\int }\frac{{\rm d}\omega ^{\prime }}{i\pi }\frac{\Gamma
_{R}[\omega ^{\prime }]}{\omega ^{\prime }-\omega -i\epsilon } 
\]
It follows that this function behaves as $\frac{1}{\omega }$ at high
frequencies 
\begin{eqnarray}
-i\omega \Gamma [\omega ] &\rightarrow &\omega _{C}\qquad {\rm for\ }\omega
\rightarrow \infty  \eqnum{16a} \\
\omega _{C} &=&{\int }\frac{{\rm d}\omega }{\pi }\Gamma _{R}[\omega ] 
\eqnum{16b}
\end{eqnarray}
Finiteness of the integral (16b) has been assumed, which requires $A>1$. The
parameter $\omega _{C}$ has the dimension of a frequency and may be
considered as the definition of the reflection cutoff if $\Gamma _{0}=1$;
otherwise, it is the product of $\Gamma _{0}$ and of the cutoff.

In order to deal with stationary fluctuations, we will require that the
mirror coupled to vacuum behaves as a stable system. Runaway solutions occur
for a perfect mirror $\omega _{C}\rightarrow \infty $, but not for a mirror
such that \cite{Position9} 
\begin{equation}
\omega _{C}\tau \ll 1  \eqnum{17}
\end{equation}
Indeed, the coupled system obeys passivity \cite{Position18} in this case.
This property is connected to the fact that vacuum cannot deliver energy for
sustaining runaway solutions, so that stability follows from passivity \cite
{Position9}. Then, a partially transmitting mirror obeying equation (1) is
stable.

\section*{Fluctuations of coupled variables}

We now come to the evaluation of coupled position and force fluctuations.

We describe the coupling by two linear response equations 
\begin{eqnarray}
F[\omega ] &=&F^{\rm in}[\omega ]+\chi _{FF}^{\rm in}[\omega ]q[\omega ]  \eqnum{18a}
\\
q[\omega ] &=&q^{\rm in}[\omega ]+\chi _{qq}^{\rm in}[\omega ]F[\omega ]  \eqnum{18b}
\end{eqnarray}
Each coupled variable, $F$ or $q$, is the sum of the corresponding input
quantity, $F^{\rm in}$ or $q^{\rm in}$, and of a linear response to the other
variable, $q$ or $F$, characterized by the linear susceptibilities $\chi
_{FF}^{\rm in}$ and $\chi _{qq}^{\rm in}$ which have been previously discussed.

The linear equations (18) are easily solved, leading to expressions of the
coupled variables in terms of the uncoupled ones and of linear
susceptibilities which are denoted $\chi _{qq}$, $\chi _{FF}$, $\chi
_{qF}=\chi _{Fq}$ for reasons which will soon become apparent 
\begin{eqnarray}
q[\omega ] &=&\chi _{qq}[\omega ]F^{\rm in}[\omega ]+\chi _{Fq}[\omega
]q^{\rm in}[\omega ]  \eqnum{19a} \\
F[\omega ] &=&\chi _{Fq}[\omega ]F^{\rm in}[\omega ]+\chi _{FF}[\omega
]q^{\rm in}[\omega ]  \eqnum{19b}
\end{eqnarray}
\begin{eqnarray}
\chi _{qq}[\omega ] &=&\frac{1}{\frac{1}{\chi _{qq}^{\rm in}[\omega ]}-\chi
_{FF}^{\rm in}[\omega ]}  \eqnum{19c} \\
\chi _{Fq}[\omega ] &=&\frac{1}{1-\chi _{FF}^{\rm in}[\omega ]\chi
_{qq}^{\rm in}[\omega ]}  \eqnum{19d} \\
\chi _{FF}[\omega ] &=&\frac{1}{\frac{1}{\chi _{FF}^{\rm in}[\omega ]}-\chi
_{qq}^{\rm in}[\omega ]}  \eqnum{19e}
\end{eqnarray}

One then deduces the commutators for the coupled operators ($\xi _{qF}=\xi
_{Fq}$; $\xi _{qF}^{\rm in}=\xi _{Fq}^{\rm in}=0$) 
\begin{eqnarray}
\xi _{qq}[\omega ] &=&\chi _{qq}[\omega ]\xi _{FF}^{\rm in}[\omega ]\chi
_{qq}[-\omega ]  \nonumber \\
&&+\chi _{Fq}[\omega ]\xi _{qq}^{\rm in}[\omega ]\chi _{Fq}[-\omega
]  \eqnum{20a} \\
\xi _{Fq}[\omega ] &=&\chi _{qq}[\omega ]\xi _{FF}^{\rm in}[\omega ]\chi
_{Fq}[-\omega ]  \nonumber \\
&&+\chi _{Fq}[\omega ]\xi _{qq}^{\rm in}[\omega ]\chi _{FF}[-\omega
]  \eqnum{20b} \\
\xi _{FF}[\omega ] &=&\chi _{Fq}[\omega ]\xi _{FF}^{\rm in}[\omega ]\chi
_{Fq}[-\omega ]  \nonumber \\
&&+\chi _{FF}[\omega ]\xi _{qq}^{\rm in}[\omega ]\chi _{FF}[-\omega
]  \eqnum{20c}
\end{eqnarray}
Using the fluctuation-dissipation relations (5,12) for the input variables,
one derives, after straighforward transformations, similar relations for the
coupled ones ($A$ and $B$ stand for $q$ or $F$) 
\begin{equation}
2i\xi _{AB}[\omega ]=\chi _{AB}[\omega ]-\chi _{AB}[-\omega ]  \eqnum{21a}
\end{equation}
One also obtains, after similar transformations 
\begin{eqnarray}
C_{AB}[\omega ] &=&2\hbar \theta (\omega )\xi _{AB}[\omega ]  \eqnum{21b} \\
\sigma _{AB}[\omega ] &=&\varepsilon (\omega )\xi _{AB}[\omega ]  \eqnum{21c}
\end{eqnarray}

These expressions constitute a consistent description of the coupled
fluctuations. The input fluctuations $q^{\rm in}$ and $F^{\rm in}$ have been treated
on an equal foot in the derivation and the resulting expressions are
symmetrical in the two quantities $q$ and $F$. The susceptibility functions 
$\chi _{qq}$, $\chi _{FF}$ and $\chi _{Fq}$ play the same role as $\chi
_{qq}^{\rm in}$ and $\chi _{FF}^{\rm in}$ and are associated to coupled variables
rather than to input ones.

We show in Appendix A that these expressions can be rewritten in a simpler,
but less symmetric, form where the input position fluctuations $q^{\rm in}$ no
longer appear. The basic reason is that the two functions $\chi _{Fq}$ and 
$\chi _{FF}$ vanish at the suspension eigenfrequencies $\pm \omega _{0}$
where the input position fluctuations $q^{\rm in}$ are concentrated. Then,
equations (19) become 
\begin{eqnarray}
q[\omega ] &=&\chi _{qq}[\omega ]F^{\rm in}[\omega ]  \eqnum{19'a} \\
F[\omega ] &=&\chi _{Fq}[\omega ]F^{\rm in}[\omega ]  \eqnum{19'b}
\end{eqnarray}
\begin{eqnarray}
\chi _{qq}[\omega ] &=&\frac{1}{m_{0}\left( \omega _{0}^{2}-\omega
^{2}\right) -\chi _{FF}^{\rm in}[\omega ]}  \eqnum{19'c} \\
\chi _{Fq}[\omega ] &=&m_{0}\left( \omega _{0}^{2}-\omega ^{2}\right) \chi
_{qq}[\omega ]  \eqnum{19'd}
\end{eqnarray}
Also, the commutators (20) are simplified to 
\begin{eqnarray}
\xi _{qq}[\omega ] &=&\chi _{qq}[\omega ]\xi _{FF}^{\rm in}[\omega ]\chi
_{qq}[-\omega ]  \nonumber \\
&=&|\chi _{qq}[\omega ]|^{2}\xi _{FF}^{\rm in}[\omega ] 
\eqnum{20'a} \\
\xi _{Fq}[\omega ] &=&\chi _{qq}[\omega ]\xi _{FF}^{\rm in}[\omega ]\chi
_{Fq}[-\omega ]  \nonumber \\
&=&m_{0}\left( \omega _{0}^{2}-\omega ^{2}\right) \xi
_{qq}[\omega ]  \eqnum{20'b} \\
\xi _{FF}[\omega ] &=&\chi _{Fq}[\omega ]\xi _{FF}^{\rm in}[\omega ]\chi
_{Fq}[-\omega ]  \nonumber \\
&=&m_{0}^{2}\left( \omega _{0}^{2}-\omega ^{2}\right) ^{2}\xi
_{qq}[\omega ]  \eqnum{20'c}
\end{eqnarray}
In other words, the position fluctuations of the coupled mirror, as well as
the coupled force fluctuations, are entirely determined by input
fluctuations $F^{\rm in}$ of vacuum radiation pressure. Although $q^{\rm in}$ and 
$F^{\rm in}$ have been treated on an equal foot, a privileged role is attributed
to $F^{\rm in}$ in the end. As usual in the theory of irreversible phenomena,
the reason for this asymmetry is that $F^{\rm in}$ corresponds to a dense
spectrum whereas $q^{\rm in}$ corresponds to a discrete one. As a result, the
linear response equations (18) can be replaced by a simple Langevin equation
(see Appendix A).

This suggests that the present approach presents some analogy with the
stochastic interpretations of quantum mechanics, like stochastic
electrodynamics \cite{Position19} or stochastic mechanics \cite{Position20}.
It indeed suggests that quantum fluctuations of mirror's position can
effectively be considered as a consequence of its coupling with zero point
fields. As a result, these fluctuations scale as $\frac{\hbar }{m_{0}}$ with 
$\hbar $ introduced through vacuum fluctuations and $m_{0}$ introduced as
the effective mass in the Newton equation of motion. It has however to be
stressed that the non-commutative character of the vacuum fluctuations plays
a central role in our derivation, whereas the zero point fields are often
treated as classical variables in stochastic interpretations.

Moreover, it has to be noted that the reverse situation could occur where
the field fluctuations would be dominated by position ones. Consider for
example the original Casimir problem \cite{Position21} with two perfect
mirrors. In this case, the intracavity field possesses a discrete spectrum.
Then, one concludes that the outside field fluctuations determine the
position fluctuations of the mirrors, turning their discrete spectrum into a
continuous one, which themselves determine the intracavity field
fluctuations.

\section*{Commutation relations for mirror's variables and effective mass}

From fluctuation-dissipation relations (21), the position commutator $\xi
_{qq}$ of the coupled mirror is directly connected to the mechanical
admittance $Y$ and impedance $Z$ (compare with the similar relation 4 for
the uncoupled mirror; $Y_{R}$ and $Z_{R}$ are the real parts, i.e. the
dissipative parts, of $Y$ and $Z$; a discussion of position fluctuations in
terms of $Y_{R}$ and $Z_{R}$ is given in Appendix A) 
\begin{eqnarray}
-i\omega \chi _{qq}[\omega ] &=&Y[\omega ]=\frac{1}{Z[\omega ]}  \eqnum{22a}
\\
\omega \xi _{qq}[\omega ] &=&Y_{R}[\omega ]=\frac{Z_{R}[\omega ]}{|Z[\omega
]|^{2}}  \eqnum{22b}
\end{eqnarray}

In the following, we rewrite the susceptibility as (see eqs 13,19'c) 
\begin{eqnarray}
\frac{1}{\chi _{qq}[\omega ]} &=&m_{0}\omega _{0}^{2}-\omega ^{2}m[\omega ] 
\eqnum{23a} \\
m[\omega ] &=&m_{0}\left( 1+i\omega \tau \Gamma [\omega ]\right)  
\eqnum{23b}
\end{eqnarray}
The coefficient $m_{0}$ appears as the quasistatic mass, that is the
effective mass at the low frequency limit. Using the relations (16) derived
from the causal properties of the susceptibility, one deduces a different
value for the high-frequency mass $m_{\infty }$ 
\begin{eqnarray}
m_{\infty } &=&m[\infty ]=m_{0}-\mu   \eqnum{23c} \\
\mu  &=&m_{0}\omega _{C}\tau =\frac{\hbar \omega _{C}}{6\pi c^{2}} 
\eqnum{23d}
\end{eqnarray}
$\mu $ appears as an `induced mass', always positive, and is similar to the
`electromagnetic mass' of classical electron theory \cite{Position8} (using
eqs 16b and 13b) 
\begin{equation}
\mu ={\int }\frac{{\rm d}\omega }{\pi }\frac{\xi _{FF}^{\rm in}[\omega ]}{\omega
^{3}}  \eqnum{23e}
\end{equation}
Here, the induced mass $\mu $ is much smaller than the quasistatic mass 
$m_{0}$ for mirrors obeying equation (1), and low- and high-frequency values
of the mass only slightly differ.

Using equations (6,22), one deduces that the canonical commutator between
velocity and position is directly connected to the admittance function, as
it is modified by coupling to vacuum (the uncoupled expression was the same
with $Y^{\rm in}$ in place of $Y$; see eq. 4) 
\begin{eqnarray}
\xi _{vq}[\omega ]=-i\omega \xi _{qq}[\omega ] &=&-\frac{i}{2}\left(
Y[\omega ]+Y[-\omega ]\right)  \eqnum{24a} \\
\left\langle \left[ v(t),q(0)\right] \right\rangle &=&-i\hbar \left(
Y(t)+Y(-t)\right)  \eqnum{24b}
\end{eqnarray}
The time dependence of the function $Y(t)$ is analysed in Appendix B.
Expanding in the small parameters $\omega _{0}\tau $ and $\omega _{C}\tau $,
assuming that $\omega _{0}$ is much smaller than $\omega _{C}$, we get the
approximate expression 
\[
Y(t)=Y^{\rm in}(t)\exp \frac{-\gamma t}{2}+\Delta Y(t) 
\]
The first term corresponds to the uncoupled admittance, which is damped on
very long times with a damping constant 
\[
\gamma \approx \omega _{0}^{2}\tau \Gamma _{0} 
\]
The second part $\Delta Y(t)$ is a `bump' centered at $t=0$, with a small
height $\frac{\omega _{C}\tau }{m_{0}}$ and a small width of the order of 
$\frac{\Gamma _{0}}{\omega _{C}}$, that is the inverse of the cutoff
frequency (see eqs 16).

The various frequencies involved in the admittance function scale as follows 
\begin{eqnarray}
\gamma  &\ll &\omega _{0}\ll \omega _{C}\ll \frac{1}{\tau }  \eqnum{25a} \\
&&\frac{\gamma }{\omega _{0}}\approx \omega _{0}\tau \Gamma _{0}  \eqnum{25b}
\end{eqnarray}
Then, the time $\frac{1}{\gamma }$ is much longer than any other
characteristic time 
\begin{equation}
\tau \ll \frac{1}{\omega _{C}}\ll \frac{1}{\omega _{0}}\ll \frac{1}{\gamma }
\eqnum{25c}
\end{equation}

One deduces (see Appendix B) that the commutator has the following values as
a function of time 
\begin{eqnarray}
\left\langle \left[ v(t),q(0)\right] \right\rangle &=&-\frac{i\hbar }{%
m_{\infty }}\qquad {\rm for\ }t\ll \frac{1}{\omega _{C}}  \eqnum{26a} \\
&=&-\frac{i\hbar }{m_{0}}\qquad {\rm for\ }\frac{1}{\omega _{C}}\ll t\ll 
\frac{1}{\omega _{0}}  \eqnum{26b} \\
&=&-\frac{i\hbar }{m_{0}}\cos \left( \omega _{0}t\right) \exp \left( -\frac{
\gamma |t|}{2}\right) \nonumber \\
&&\qquad {\rm for\ }\frac{1}{\omega _{C}}\ll t 
\eqnum{26c}
\end{eqnarray}
The commutator at times shorter than the oscillation period, in particular,
the commutator for an unbound mirror $\omega _{0}=0$, has a time dependence
directly connected to the difference between the two masses which appear in
equations (23). At short times $t\ll \frac{1}{\omega _{C}}$ and longer times 
$\frac{1}{\omega _{C}}\ll t$ respectively, the commutator is related to the
high frequency mass $m_{\infty }$ (mass for $\omega _C \ll \omega $)
and to the quasistatic mass $m_{0}$ (mass for $\omega \ll \omega _C $).

This discussion suggests that $m_{\infty }$ has to be considered as the bare
mass while $m_{0}$ contains a part due to coupling with vacuum. In this
spirit, we may rewrite equations (23) as 
\begin{eqnarray}
m[\omega ] &=&m_{\infty }+\mu [\omega ]  \eqnum{23'a} \\
\mu [\omega ] &=&m_{0}\tau \left( \omega _{C}+i\omega \Gamma [\omega ]\right)
\eqnum{23'b} \\
\mu [0] &=&m_{0}\omega _{C}\tau =\mu \qquad \mu [\infty ]=0  \eqnum{23'c}
\end{eqnarray}
where the frequency dependent function $\mu [\omega ]$ appears as a
contribution of the vacuum energy bound to the mirror, and swept along by
its motion. This phenomenon is effective at low frequencies, but not at high
frequencies since the mirror is transparent at this limit and its motion is
decoupled from vacuum fields.

\section*{Position and velocity variances}

We now come to the evaluation of the position and velocity variances which
can be deduced as integrals of the symmetrical correlation function (see eqs
9).

Before evaluating these integrals, we derive a Schwartz inequality for the
product of the two variances 
\begin{eqnarray*}
\overline{\left( 1+\alpha |\omega |\right) ^{2}\sigma _{qq}} &\geq &0\qquad 
{\rm for\ any}\ \alpha \ {\rm real} \\
\overline{\sigma _{qq}}\overline{\omega ^{2}\sigma _{qq}} &\geq &
\left( \overline{ |\omega |\sigma _{qq}}\right)^{2}
\end{eqnarray*}
Using equations (21c) and (24), one gets 
\[
\sigma _{qq}[\omega ]=\varepsilon (\omega )\xi _{qq}[\omega ]\qquad 
\overline{|\omega |\sigma _{qq}}=\overline{\omega \xi _{qq}}=i\xi _{vq}(0)
\]
leading to a Heisenberg inequality, which sets a lower bound on the product
of variances, related to the equal-time canonical commutator 
\begin{equation}
\Delta q\Delta v\geq \frac{\hbar }{2m_{\infty }}\geq \frac{\hbar }{2m_{0}} 
\eqnum{27}
\end{equation}

In order to explicitly evaluate the variances, we write the noise spectra as
(see eqs 20') 
\begin{eqnarray}
\xi _{qq}[\omega ] &=&\frac{1}{m_{0}}\frac{\omega ^{3}\tau \Gamma _{R}}{
\left( \omega ^{2}-\omega _{0}^{2}-\omega ^{3}\tau \Gamma _{I}\right)
^{2}+\left( \omega ^{3}\tau \Gamma _{R}\right) ^{2}}  \eqnum{28a} \\
\sigma _{qq}[\omega ] &=&\varepsilon (\omega )\xi _{qq}[\omega ]=|\xi
_{qq}[\omega ]|  \eqnum{28b}
\end{eqnarray}
As $\tau $ is a very small time, the noise spectrum $\sigma _{qq}$ is
approximately a sum of two narrow peaks around the two suspension
eigenfrequencies. The width of the peaks is the damping constant $\gamma $
encountered previously in the time dependence of the commutator. Besides
these peaks, the noise spectrum also contains a small background (see
Appendix B). If one considers the limit where the mirror is decoupled from
vacuum, one gets a vanishing width $\gamma $ for the peaks and a vanishing
height for the background (compare with the discussion of the time
dependence of $Y(t)$). Then, a simple expression is obtained, which
corresponds to the uncoupled oscillator (see eqs 3), so that the usual
variances (11) are recovered. However, this limit must be considered more
carefully, as shown by an examination of the convergence of the integrals
(9).

Due to the high frequency behaviour of $\Gamma _{R}[\omega ]$ (see eq. 15),
a finite value is obtained for $\Delta q^{2}$ as soon as $A>0$ ($A>1$ was
already needed for getting a finite value for $\omega _{C}$). It can be
checked that the value for $\Delta q^{2}$ is close to the expression (11)
associated with an uncoupled oscillator ($\tau \rightarrow 0$ or $\Gamma
_{0}\rightarrow 0$). A more stringent condition $A>2$ is required in order
to get a finite value for $\Delta v^{2}$; for $A=2$, a logarithmic
divergence is obtained. This implies that the velocity variance may be
completely changed by the coupling, even for a vanishing value of $\tau $ or 
$\Gamma _{0}$. The fact that we may obtain an infinite variance for the
instantaneous velocity is not surprising, a similar result being obtained in
classical theory of Brownian motion.

In the particular case of an unbound mirror ($\omega _{0}=0$) coupled to
vacuum, one gets 
\begin{eqnarray}
\xi _{qq}[\omega ] &=&\frac{1}{m_{0}\omega }\frac{\tau \Gamma _{R}}{\left(
1-\omega \tau \Gamma _{I}\right) ^{2}+\left( \omega \tau \Gamma _{R}\right)
^{2}}  \eqnum{29a} \\
\sigma _{qq}[\omega ] &=&|\xi _{qq}[\omega ]|  \eqnum{29b}
\end{eqnarray}
The low frequency divergence of the spectra (29) leads to an infinite
position variance $\Delta q^{2}$, typical of the unbound diffusion process
experienced by a free particle submitted to a random force. For a mirror in
vacuum, it could be expected that the stationary state is, as in standard
quantum mechanics, a velocity eigenstate ($\Delta v^{2}=0$; $\Delta
q^{2}=\infty $). However, the integral $\Delta v^{2}$ diverges if $A\leq 2$,
as in the general case of a suspended mirror's motion.

\section*{Quantum diffusion of a mirror coupled to vacuum}

The quantum diffusion of a mirror coupled to vacuum may be characterized
quantitatively by the function 
\begin{eqnarray*}
\Delta (t)&=&\frac{1}{2}\left\langle \left( q(t)-q(0)\right) ^{2}\right\rangle
=\hbar \left( \sigma _{qq}(0)-\sigma _{qq}(t)\right) \\
&=&{\int }\frac{{\rm d}
\omega }{2\pi }\hbar \sigma _{qq}[\omega ]\left( 1-\cos \left( \omega
t\right) \right) 
\end{eqnarray*}
This integral is well defined for any value of $t$, even for an unbound
mirror ($\omega _{0}=0$), and it vanishes when $t$ goes to zero. This is
also the case for the integral associated with $\Delta ^{\prime }(t)$ (we
suppose $A>1$), but not for $\Delta ^{\prime \prime }(t)$, which is not
defined if $A\leq 2$ ($\Delta ^{\prime \prime }(0)=\Delta v^{2}$).

Interesting results about quantum diffusion are obtained by using the
analytic properties of the correlation function $C_{qq}(t)$. Since the noise
spectrum $C_{qq}[\omega ]$ contains only positive frequency components (see
eq. 21b), $C_{qq}(t)$ is analytic and regular in the half plane $\Im t<0$
\cite{Position22,Position23}. Then, $\sigma _{qq}(t)$ may be deduced from 
$\xi _{qq}(t)$ through a dispersion relation (${\cal P}$ stands for a
principal value) 
\begin{equation}
\sigma _{qq}(t)={\int }{\rm d}t^{\prime }\xi _{qq}(t-t^{\prime }){\cal P}
\frac{1}{i\pi t^{\prime }}  \eqnum{30}
\end{equation}
One obtains a simple integral relation between the time dependent commutator
and the position variance by setting $t=0$ in the preceding integral. One
also obtains by deriving the expression of $\sigma _{qq}(t)$ 
\[
\sigma _{qq}^{\prime }(t)={\int }{\rm d}t^{\prime }\xi _{qq}^{\prime
}(t-t^{\prime }){\cal P}\frac{1}{i\pi t^{\prime }}
\]

In the particular case of an unbound mirror, $\xi _{qq}^{\prime }$ may be
written (see Appendix B) 
\[
\xi _{qq}^{\prime }(t)=-i\left( \frac{1}{2m_{0}}+\Delta Y_{R}(t)\right) 
\]
The contribution to the integral $\sigma _{qq}^{\prime }(t)$ of the first
term, which is time independent, vanishes. Since $\Delta Y_{R}$ differs from
zero only at short times, of the order of $\frac{1}{\omega _{C}}$, the
asymptotic behaviour at long times of $\sigma _{qq}^{\prime }(t)$ can be
evaluated as 
\[
\sigma _{qq}^{\prime }(t)\approx -\frac{\Delta Y_{R}[0]}{\pi t}\approx -%
\frac{\tau \Gamma _{0}}{\pi m_{0}t}\qquad {\rm for\ }\omega _{C}t\gg 1
\]
It follows that the quantum diffusion of an unbound mirror coupled to vacuum
is characterized by a logarithmic behaviour at long times \cite{Position24} 
\[
\Delta (t)\approx \frac{\hbar \tau \Gamma _{0}}{\pi m_{0}}\ln \frac{t}{t_{0}}
\qquad {\rm for\ }\omega _{C}t\gg 1
\]
$t_{0}$ is an integration constant, of the order of $\frac{1}{\omega _{C}}$.
This diffusion corresponds to very small displacements, the length scale
being in fact the Compton wavelength $\frac{\hbar }{m_{0}c}$ associated with
the mirror 
\begin{equation}
\Delta (t)\approx \frac{\Gamma _{0}}{6\pi ^{2}}\left( \frac{\hbar }{m_{0}c}
\right) ^{2}\ln \left( \omega _{C}t\right)   \eqnum{31}
\end{equation}
It has to be noted that thermal effects \cite{Position15} change the
behaviour of $\Delta (t)$ at long times, even at a low temperature.

\section*{Fluctuations of coupled force}

It is interesting to discuss the autocorrelation of the coupled force $F$,
that is the force computed in presence of the mirror's radiation reaction,
and its correlation with the coupled position.

The autocorrelation of the coupled force $F$ is given by equation (20'c); 
$\xi _{FF}$ is identical to $\xi _{FF}^{\rm in}$, except in the narrow resonance
peaks where the fluctuations of the coupled force are smaller than those of
the input force (compare with eqs 28) 
\begin{equation}
\xi _{FF}[\omega ]=\xi _{FF}^{\rm in}[\omega ]\frac{\left( \omega ^{2}-\omega
_{0}^{2}\right) ^{2}}{\left( \omega ^{2}-\omega _{0}^{2}-\omega ^{3}\tau
\Gamma _{I}\right) ^{2}+\left( \omega ^{3}\tau \Gamma _{R}\right) ^{2}} 
\eqnum{32}
\end{equation}

The correlation of the coupled force with the coupled position is described
by equation (20'b). This expression implies that coupled position and
coupled force evaluated at equal times commute; $\xi _{Fq}[\omega ]$ is an
odd function of $\omega $ as $\xi _{qq}[\omega ]$, so that 
\begin{equation}
\xi _{Fq}(0)=0  \eqnum{33}
\end{equation}
From the same argument of parity, it follows that coupled velocity and
coupled force do not necessarily commute when evaluated at equal time.

One also deduces the symmetrical correlation function between $q$ and $F$ 
\begin{eqnarray}
\sigma _{Fq}[\omega ] &=&\varepsilon (\omega )\xi _{Fq}[\omega ]=m_{0}\left(
\omega _{0}^{2}-\omega ^{2}\right) \sigma _{qq}[\omega ]  \nonumber \\
\left\langle q(0)F(0)\right\rangle &=&\left\langle F(0)q(0)\right\rangle
=\hbar \overline{\sigma _{Fq}}  \nonumber \\
&=&m_{0}\left( \omega _{0}^{2}\Delta
q^{2}-\Delta v^{2}\right)  \eqnum{34}
\end{eqnarray}
which is reminiscent of the virial theorem \cite{Position25}.

It appears from this discussion that the position of the mirror and the
vacuum pressure exerted upon it are intimately correlated: at each
frequency, they are directly proportional to each other. Therefore the
quantum fluctuations of the mirror cannot be treated independently of those
of vacuum radiation pressure.

\section*{Ultimate quantum limits in a position measurement}

We show in this section that the noise spectrum $\sigma _{qq}$ can be
considered as the ultimate quantum limit for the sensitivity in a position
measurement, when mirrors coupled to vacuum are used. More precisely, we
discuss how this limit may be reached, at least in principle; we will not
analyse in detail the effective realization of this sensitivity.

We consider a simple model of position measurement where a laser is sent
onto the mirror and where the phase $\Phi $ of the reflected beam is
monitored \cite{Position3}. This phase $\Phi $ is related to the mirror's
position $q$ through ($k_{0}$ is the laser wavenumber) 
\[
\Phi (t)=2k_{0}\left( q(t)+\delta q_{RP}(t)\right) +\delta \Phi (t) 
\]
Here, $q(t)$ represents the position discussed in foregoing sections, with
its quantum fluctuations; $\delta \Phi $ represents the quantum fluctuations
of the phase of the incident field, and $\delta q_{RP}$ stands for the
random motion of the mirror due to the radiation pressure $\delta F_{RP}$
exerted upon it \cite{Position1}, proportional to the fluctuations $\delta I$
of laser intensity 
\[
\delta F_{RP}(t)=2\hbar k_{0}\delta I(t) 
\]
In the following, we will consider that the fluctuations of $q$, which can
be assigned to vacuum radiation pressure (the vacuum force at frequency 
$\omega $ is associated with vacuum field fluctuations at frequencies
comprised between 0 and $\omega $), are not correlated with the fluctuations
of $\delta q_{RP}$ and $\delta \Phi $ (field fluctuations at frequencies
close to the laser frequency). We will also assume that the field
fluctuations around the laser frequency can be transformed (squeezed) at our
convenience, while the lower frequencies vacuum fluctuations are not
modified.

In a linear analysis, the response of the mirror to radiation pressure is
described by the susceptibility function 
\[
\delta q_{RP}[\omega ]=\chi _{qq}[\omega ]\delta F_{RP}[\omega ] 
\]
The estimator $\widehat{q}[\omega ]$ for a frequency component $q[\omega ]$
of position can be written 
\[
\widehat{q}[\omega ]=\frac{\Phi [\omega ]}{2k_{0}}=q[\omega ]+q_{N}[\omega ] 
\]
The added noise $q_{N}[\omega ]$ is a sum of two terms associated
respectively with phase and intensity fluctuations (compare with eqs 11 of
ref. \cite{Position3}) 
\[
q_{N}[\omega ]=\frac{\delta \Phi [\omega ]}{2k_{0}}+2\hbar k_{0}\chi
_{qq}[\omega ]\delta I[\omega ] 
\]

When phase and intensity fluctuations are considered as uncorrelated, the
corresponding noise spectrum is obtained as a sum of two contributions \cite
{Position1} 
\[
\sigma _{q_{N}q_{N}}[\omega ]=\frac{\sigma _{\Phi \Phi }[\omega ]}{\left(
2k_{0}\right) ^{2}}+\left( 2\hbar k_{0}\right) ^{2}|\chi _{qq}[\omega
]|^{2}\sigma _{II}[\omega ]
\]
The minimum value reached by this expression when the various parameters are
varied, with variations of $\sigma _{\Phi \Phi }$ and $\sigma _{II}$
constrained by a Heisenberg inequality, is the standard quantum limit 
\[
\sigma _{q_{N}q_{N}}^{{\rm SQL}}[\omega ]=|\chi _{qq}[\omega ]|
\]

It is known that the sensitivity can be pushed beyond the standard quantum
limit \cite{Position2}. Indeed, phase and intensity fluctuations are
linearly superimposed in the fluctuations of the monitored signal, which
therefore depend also upon the correlation $\sigma _{I\Phi }$. Consequently,
one can reduce noise by squeezing a well chosen quadrature component of the
fields. A lower bound $\sigma ^{{\rm UQL}}$ is obtained by optimising the
parameters characterizing the squeezed fields, the variations of $\sigma
_{\Phi \Phi }$, $\sigma _{II}$ and $\sigma _{I\Phi }$ being constrained by a
Heisenberg inequality. It is determined by the dissipative part of the
mirror's susceptibility \cite{Position3}. With notations of the present
paper, this `ultimate quantum limit' can be written 
\[
\sigma _{q_{N}q_{N}}^{{\rm UQL}}[\omega ]=|\xi _{qq}[\omega ]|=\sigma
_{qq}[\omega ]
\]
We have used relation (28b) between the noise spectrum and the dissipative
part of the susceptibility.

It is instructive to measure the fluctuations of the estimated position 
$\widehat{q}$ as a noise energy $N$ per unit bandwidth (fluctuations of $q$
and $q_{N}$ are uncorrelated) 
\[
N[\omega ]=\frac{m_{0}}{2}\left( \omega _{0}^{2}+\omega ^{2}\right) \left(
C_{qq}[\omega ]+C_{q_{N}q_{N}}[\omega ]\right) 
\]
The standard quantum limit corresponds to 
\begin{eqnarray}
N^{{\rm SQL}}[\omega ] &=&\hbar \theta (\omega )m_{0}  \times \nonumber \\
&&\left( \omega
_{0}^{2}+\omega ^{2}\right) \left( |\xi _{qq}[\omega ]|+|\chi _{qq}[\omega
]|\right)   \eqnum{35a} \\
&\approx &\hbar \theta (\omega )\qquad {\rm for\ }\omega _{0}\ll \omega \ll
\omega _{S}  \eqnum{35b}
\end{eqnarray}
This limit is entirely determined by added fluctuations (second term in 35a)
and proper quantum fluctuations of position (first term in 35a) do not
contribute (these fluctuations are concentrated in a very narrow peak at the
suspension eigenfrequency). This confirms that proper fluctuations can be
forgotten when analyzing a position measurement with a sensitivity level
around standard quantum limit.

In contrast, the ultimate quantum limit reaches the level set by the
dissipative character of coupling with vacuum; proper fluctuations have the
same contribution as added fluctuations 
\begin{eqnarray}
N^{{\rm UQL}}[\omega ] &=&\hbar \theta (\omega )m_{0}\left( \omega
_{0}^{2}+\omega ^{2}\right) 2|\xi _{qq}[\omega ]|  \eqnum{36a} \\
&\approx &2\hbar \theta (\omega )\omega \tau \Gamma _{0}\qquad {\rm for\ }
\omega _{0}\ll \omega \ll \omega _{S}  \eqnum{36b}
\end{eqnarray}
This limit (36) is far beyond the standard quantum limit (35). It has to be
stressed however that the input state of the field used in the measurement
must be carefully controlled, in the frequency band around laser frequency
which is involved in the measurement. Reaching effectively the ultimate
limit in a real experiment may be a rather difficult task.

This is made clear by expressing this limit as a position variance 
\begin{eqnarray*}
\Delta \widehat{q}^{2}\approx &&\frac{2B}{\omega }\frac{2\Gamma _{0}}{3\pi }
\left( \frac{\hbar }{m_{0}c}\right) ^{2}  \\
&&\qquad {\rm for\ }\omega _{0}\ll \omega -B,\omega + B \ll \omega _{S} 
\end{eqnarray*}
where $\frac{\hbar }{m_{0}c}$ is the Compton wavelength of the mirror and 
$2B $ the measurement bandwidth (signal measured through a filter
characterized by a function $G[\omega ]$ having a maximum value of 1 at
expected signal frequency) 
\[
2B=\overline{G}
\]
The effective noise $\Delta \widehat{q}^{2}$ in the measurement may be much
smaller than the position variance: the integral of a positive quantity over
a limited bandwidth is obviously smaller than the integral of the same
quantity over all frequencies (see eqs 9). This occurs, if the input state
of the laser is optimized, when the suspension eigenfrequencies are outside
the detection bandwidth; in the opposite case, the effective noise would be
of the order of the position variance. Incidentally, this discussion means
that quantum noise may be reduced by a narrow band detection, exactly in the
same manner as thermal noise \cite{Position26}.

\section*{Discussion}

Any mirror, more generally any scatterer, is coupled to vacuum radiation
pressure, so that its position acquires quantum fluctuations from vacuum.
This mechanism is quite analogous to the thermalization of the variables of
a system coupled to a thermal bath, but it involves vacuum (i.e. zero
temperature) fluctuations.

Quite generally, coupling a classical quantity to quantum ones leads to
inconsistencies \cite{Position27}. In contrast, generic quantum properties
are preserved in a consistent treatment, as a consequence of
fluctuation-dissipation relations. Then, position fluctuations of the
coupled mirror are connected to the dissipative part of its mechanical
admittance.

It appears that fluctuations of the coupled variables are determined by
input fluctuations corresponding to a continuous spectrum (i.e. the
`dissipative' ones), which are here the force fluctuations. It can therefore
be asserted that quantum properties of position are generated from vacuum
fluctuations: the same results would be obtained if the position were
initially treated as a classical number. Nevertheless, the proper position
fluctuations are recovered in the end, in the limit of decoupling. Coupling
with vacuum is very weak, making the value of $\tau $ extremely small, and
this justifies that usual quantum mechanics can provide us with a good
description of position fluctuations.

The canonical commutation relation between position and velocity is slightly
modified, this modification being only effective at small times of the order
of the reflection delay. This time dependence is connected to the difference
between the bare mass (at high frequencies where field energy cannot follow
the mirror's motion) and the quasistatic mass (at low frequencies where a
finite part of the vacuum energy is swept along the mirror's motion).

The noise spectrum, which describes the quantum fluctuations of position for
a mirror coupled to vacuum, consists in two narrow resonance peaks close to
the suspension eigenfrequencies and a small broad background. When the peaks
are approximated as Dirac distributions and the background disregarded,
integrals over frequency of the noise spectrum reproduce the dispersions 
$\Delta q$ and $\Delta v$ expected for a quantum harmonic oscillator in its
ground state.

However, differences exist which may have some importance. Although finite
in the uncoupled case, the velocity dispersion may be infinite for the
coupled oscillator, a property reminiscent of classical theory of Brownian
motion. Also, the resonance peaks have a small width, typical of dissipative
coupling with vacuum, so that the correlation functions are damped on very
long times and differ from the usual dissipationless expressions. For an
unbound mirror coupled to vacuum, a logarithmic diffusion is obtained (no
diffusion for an unbound uncoupled mirror), the length scale of which is
given by the Compton wavelength associated with the mirror.

Finally, the proper quantum fluctuations of mirrors used in a length
measurement such as interferometric detection of gravitational waves are
accounted for as soon as coupling with vacuum radiation pressure is treated.
The noise spectrum associated with position sets the ultimate bound on
sensitivity when the measurement is optimized. Then, the effective noise in
the measurement can be much smaller than the position variance computed from
quantum mechanics, as soon as the detection bandwidth does not contain the
suspension eigenfrequencies. Quantum noise, as other sources of noise, may
be reduced by monitoring a signal away from resonance frequencies.

\medskip \noindent {\bf Acknowlegdements}

Thanks are due to J.M. Courty, A. Heidmann, P.A. Maia Neto and J. Maillard
for discussions.

\appendix 

\section{Elimination of the input position fluctuations}

The linear response equations (18) can be replaced by simpler, but less
symmetric, ones where the input position fluctuations $q^{\rm in}$ no longer
appear.

The demonstration is based upon the following relations (deduced from eqs
19) 
\begin{eqnarray*}
\chi _{Fq}[\omega ] &=&\frac{\chi _{qq}[\omega ]}{\chi _{qq}^{\rm in}[\omega ]}
=m_{0}\left( \omega _{0}^{2}-\omega ^{2}\right) \chi _{qq}[\omega ] \\
\chi _{FF}[\omega ] &=&\chi _{Fq}[\omega ]\chi _{FF}^{\rm in}[\omega
]=m_{0}\left( \omega _{0}^{2}-\omega ^{2}\right) \chi _{qq}[\omega ]\chi
_{FF}^{\rm in}[\omega ]
\end{eqnarray*}
It follows that the functions $\chi _{Fq}$ and $\chi _{FF}$ vanish at the
suspension eigenfrequencies 
\[
\chi _{Fq}[\pm \omega _{0}]=\chi _{FF}[\pm \omega _{0}]=0
\]
Since the input position fluctuations $q^{\rm in}$ are concentrated at these
frequencies (see eqs 3), they cannot feed the fluctuations of the coupled
variables which are therefore entirely determined by the input force
fluctuations. Equations (19) become 
\begin{eqnarray*}
q[\omega ] &=&\chi _{qq}[\omega ]F^{\rm in}[\omega ] \\
F[\omega ] &=&m_{0}\left( \omega _{0}^{2}-\omega ^{2}\right) q[\omega ] \\
\chi _{qq}[\omega ] &=&\frac{1}{m_{0}\left( \omega _{0}^{2}-\omega
^{2}\right) -\chi _{FF}^{\rm in}[\omega ]}
\end{eqnarray*}
Although $q^{\rm in}$ and $F^{\rm in}$ have been treated on an equal foot, $F^{\rm in}$
plays a dominant role in the end. This is due to the property 
\[
\frac{\xi _{qq}^{\rm in}[\omega ]}{\chi _{qq}^{\rm in}[\omega ]}=0
\]
which is satisfied for the discrete spectrum of equations (3). The same
property is not fulfilled by the dense spectrum associated with the force
(see eq. 13). It results that expressions (20) of the commutators can be
rewritten as (20').

A mirror coupled to vacuum then obeys a simple equation of motion where, as
in Langevin theory of Brownian motion \cite{Position14}, the force $F$ is a
sum of a Langevin force $F^{\rm in}$, the input force, and of a long term
cumulative effect, the motional force, 
\begin{eqnarray*}
&&m_{0}\omega _{0}^{2}q(t)+m_{0}q^{\prime \prime }(t)=F(t) \\
&&m_{0}\left( \omega _{0}^{2}-\omega ^{2}\right) q[\omega ]=F[\omega ] \\
&&F[\omega ]=F^{\rm in}[\omega ]+\chi _{FF}^{\rm in}[\omega ]q[\omega ]
\end{eqnarray*}
These equations can be obtained directly by setting $q^{\rm in}=0$ in (18).

These equations can be solved in terms of a mechanical impedance $Z$ or of a
mechanical admittance $Y$, which connect the applied force and the effective
velocity 
\begin{eqnarray*}
&&F^{\rm in}[\omega ] = Z[\omega ]v[\omega ] \\
&&Z[\omega ] = \frac{m_{0}\omega_{0}^{2}}{-i\omega }-i\omega m_{0}
\left( 1+i\omega \tau \Gamma [\omega]\right) \\
&&v[\omega ] =Y[\omega ]F^{\rm in}[\omega ]\\
&&Y[\omega ]=\frac{1}{Z[\omega ]}=-i\omega \chi _{qq}[\omega ]
\end{eqnarray*}
The dissipative parts of the impedance and admittance functions, defined as
their real parts $Z_{R}$ and $Y_{R}$ in the frequency domain, are positive
at all frequencies (see eq 14) 
\[
Z_{R}[\omega ]=m_{0}\omega ^{2}\tau \Gamma _{R}[\omega ]\geq 0\qquad
Y_{R}[\omega ]=\frac{Z_{R}[\omega ]}{|Z[\omega ]|^{2}}\geq 0 
\]
In fact, the impedance and admittance functions are passive functions, which
ensures stability of the coupled system \cite{Position9}.

We can then give a simpler derivation, in the spirit of Nyquist's derivation 
\cite{Position11,Position12,Position13}, of the fluctuation-dissipation
relations for coupled position. We first rewrite the relation between the
commutator $\xi _{FF}^{\rm in}$ and the dissipative part of the motional
susceptibility $\chi _{FF}^{\rm in}$ (see eqs 12) 
\[
\xi _{FF}^{\rm in}[\omega ]=\omega Z_{R}[\omega ] 
\]
Hence, the non commutative character of velocity fluctuations is related to
the dissipative part $Y_{R}$ of the mechanical admittance 
\[
\xi _{vv}[\omega ]=|Y[\omega ]|^{2}\xi _{FF}^{\rm in}[\omega ]=\omega
Y_{R}[\omega ] 
\]
The result for the position's commutator is the same as previously obtained
(see eqs 21) 
\[
\xi _{qq}[\omega ]=\frac{\xi _{vv}[\omega ]}{\omega ^{2}}=\frac{Y_{R}[\omega
]}{\omega } 
\]

\section{Time dependence of the commutator}

In order to analyze the time dependence of the coupled commutator $\xi
_{vq}[\omega ]$, we first consider the case of an unbound mirror coupled to
vacuum, whose admittance can be written ($\omega _{0}=0$) 
\begin{eqnarray*}
Y[\omega ] &=&\frac{1}{\left( \epsilon -i\omega \right) m[\omega ]}=\frac{1}
{m_{0}\left( \epsilon -i\omega \right) }+\Delta Y[\omega ] \\
\Delta Y[\omega ] &=&\frac{\tau \Gamma [\omega ]}{m_{0}\left( 1+i\omega \tau
\Gamma [\omega ]\right) } \\
Y(t) &=&\frac{\theta (t)}{m_{0}}+\Delta Y(t)
\end{eqnarray*}
In this decomposition, the first part is the uncoupled admittance while the
second part represents a correction.

One deduces $Y(0^{+})$ from the high frequency behaviour of $Y[\omega ]$ 
\[
Y(0^{+})=\lim_{\omega \rightarrow \infty }\left( -i\omega Y[\omega ]\right) =
\frac{1}{m_{\infty }} 
\]
Starting from the value $\frac{1}{m_{\infty }}$ for $t=0^{+}$, $Y(t)$
decreases to the slightly different value $\frac{1}{m_{0}}$ for $%
t\rightarrow \infty $. In other words, $Y(t)$ is the sum of a constant value 
$\frac{1}{m_{0}}$ and of a bump $\Delta Y(t)$, having a height 
\[
\Delta Y(0^{+})=\frac{1}{m_{\infty }}-\frac{1}{m_{0}}=\frac{1}{m_{0}}\frac{
\omega _{C}\tau }{1-\omega _{C}\tau }\approx \frac{\omega _{C}\tau }{m_{0}} 
\]
The integral of the bump is given by 
\[
\Delta Y[0]=\frac{\tau \Gamma _{0}}{m_{0}} 
\]
so that the width of the bump is of the order of $\frac{\Gamma _{0}}{\omega
_{C}}$.

One deduces from equations (24) 
\[
\left\langle \left[ v(t),q(0)\right] \right\rangle =-\frac{i\hbar }{m_{0}}
-i\hbar \left( \Delta Y(t)+\Delta Y(-t)\right) 
\]
The first term is a canonical commutator (see eq. 10) while the second is a
bump-shaped correction, having a width of the order of the mean reflexion
delay $\frac{1}{\omega _{C}}$. At short and long times, one finds equations
(26a,b) 
\begin{eqnarray*}
\left\langle \left[ v(t),q(0)\right] \right\rangle  &=&-\frac{i\hbar }
{m_{\infty }}\qquad {\rm for\ }t\ll \frac{1}{\omega _{C}} \\
&=&-\frac{i\hbar }{m_{0}}\qquad {\rm for\ }\frac{1}{\omega _{C}}\ll t
\end{eqnarray*}

For an harmonically suspended mirror, the admittance function $Y$ can be
written as a sum of three components rather than two 
\begin{eqnarray*}
Y[\omega ] &=&\frac{\rho _{+}}{i\omega _{+}-i\omega }+\frac{\rho _{-}}
{i\omega _{-}-i\omega }+\Delta Y[\omega ] \\
Y(t) &=&\theta (t)\left( \rho _{+}\exp (-i\omega _{+}t)+\rho _{-}\exp
(-i\omega _{-}t)\right) +\Delta Y(t)
\end{eqnarray*}
The two complex numbers $\omega _{\pm }$ are the poles of the admittance
function $Y[\omega ]$, which are close to the suspension eigenfrequencies 
$\pm \omega _{0}$ 
\begin{eqnarray*}
\omega _{\pm } &=&\pm \overline{\omega }_{0}-\frac{i\gamma }{2}\qquad
m_{0}\omega _{0}^{2}=\omega _{\pm }^{2}m[\omega _{\pm }] \\
\rho _{\pm } &=&\frac{1}{2m[\omega _{\pm }]+\omega _{\pm }m^{\prime }[\omega
_{\pm }]}
\end{eqnarray*}
We can estimate these quantities in an expansion with respect to the small
parameters $\omega _{0}\tau $ and $\omega _{C}\tau $. Assuming that $\omega
_{0}$ is much smaller than $\omega _{C}$, we get at lowest order 
\begin{eqnarray*}
\gamma  &\approx &\omega _{0}^{2}\tau \Gamma _{0} \\
\overline{\omega }_{0}^{2} &\approx &\omega _{0}^{2}+\frac{\gamma ^{2}}{4} \\
\rho _{\pm } &\approx &\frac{1}{m_{0}\left( 2+3i\omega _{\pm }\tau \Gamma
_{0}\right) }
\end{eqnarray*}
One then deduces by straightforward calculations 
\[
\Delta Y[0]\approx \frac{\tau \Gamma _{0}}{m_{0}}\qquad \Delta
Y(0^{+})\approx \frac{\omega _{C}\tau }{m_{0}}
\]
so that the shape of the function $\Delta Y$ is similar to that previously
obtained for an unbound mirror.

Consequently, at short and long times, one finds (26a,c) 
\begin{eqnarray*}
\left\langle \left[ v(t),q(0)\right] \right\rangle &=&-\frac{i\hbar }
{m_{\infty }}\qquad {\rm for\ }t\ll \frac{1}{\omega _{C}} \\
&=&-\frac{i\hbar }{m_{0}}\cos \left( \overline{\omega }_{0}t\right) \exp
\left( -\frac{\gamma |t|}{2}\right) \\
&&\qquad {\rm for\ }\frac{1}{\omega _{C}}
\ll t
\end{eqnarray*}

\section{Case of an anharmonic suspension}

For the sake of simplicity, the simple case of an harmonically suspended
mirror has been considered throughout the paper. We show here how the
results can be generalized to the case of an anharmonic suspension.

Denoting $\omega _{\alpha }$ the various excitation frequencies from the
ground state to the upper states, and $q_{\alpha }$ the corresponding matrix
elements of the position operator, one obtains 
\[
C_{qq}^{\rm in}(t)=\sum_{\alpha }|q_{\alpha }|^{2}\exp (-i\omega _{\alpha }t) 
\]
Hence 
\begin{eqnarray*}
\xi _{qq}^{\rm in}[\omega ] &=&\frac{1}{2\hbar }\sum_{\alpha }|q_{\alpha
}|^{2}2\pi \left( \delta (\omega -\omega _{\alpha })-\delta (\omega +\omega
_{\alpha })\right) \\
C_{qq}^{\rm in}[\omega ] &=&\sum_{\alpha }|q_{\alpha }|^{2}2\pi \delta (\omega
-\omega _{\alpha })=2\hbar \theta (\omega )\xi _{qq}^{\rm in}[\omega ] \\
\sigma _{qq}^{\rm in}[\omega ] &=&\frac{1}{2\hbar }\sum_{\alpha }|q_{\alpha
}|^{2}2\pi \left( \delta (\omega -\omega _{\alpha })+\delta (\omega +\omega
_{\alpha })\right) \\
&=&\varepsilon (\omega )\xi _{qq}^{\rm in}[\omega ]
\end{eqnarray*}
One also defines the susceptibility 
\begin{eqnarray*}
&&\chi _{qq}^{\rm in}[\omega ]=\frac{1}{\hbar }\sum_{\alpha }|q_{\alpha }|^{2}
\frac{2\omega _{\alpha }}{\omega _{\alpha }^{2}+(\epsilon -i\omega )^{2}} \\
&&\chi _{qq}^{\rm in}[\omega ]-\chi _{qq}^{\rm in}[-\omega ]=2i\xi _{qq}^{\rm in}[\omega
]
\end{eqnarray*}

The frequency dependence of the susceptibility is less simple than in the
harmonic case. However, the fluctuation-dissipation relations remain valid,
for coupled as well as uncoupled variables. Also, the following property
remains true for a discrete spectrum 
\begin{eqnarray*}
\frac{1}{\chi _{qq}^{\rm in}[\pm \omega _{\alpha }]} &=&0 \\
\frac{\xi _{qq}^{\rm in}[\omega ]}{\chi _{qq}^{\rm in}[\omega ]} &=&0
\end{eqnarray*}
which allows one to express the coupled fluctuations in terms of the
uncoupled force fluctuations only.

\end{document}